\documentclass[a4paper,twoside,12pt]{article}

\usepackage{adm}

\begin{document}

\title[Реализация параллельного алгоритма поиска кратчайшего вектора в блочном методе Коркина-Золотарева]
      {Реализация параллельного алгоритма поиска кратчайшего вектора в блочном методе Коркина-Золотарева}%
      {Реализация параллельного алгоритма поиска кратчайшего вектора в блочном методе Коркина-Золотарева}
      {The implementation of the parallel shortest vector enumerate in the block Korkin-Zolotarev method }
\author{В.\,С.~Усатюк}%
       {Усатюк~В.\,С.}
       {Usatyuk~V.\,S.}
\email{L@Lcrypto.com} 
\organization{Братский государственный университет, г.~Братск, Россия}
\udk{511.9}
\maketitle

\begin{abstract}
 В работе предложена параллельная реализация алгоритма Каннана для решения задач поиска кратчайшего и короткого векторов в решетке. Алгоритм может применяться в составе блочного метода Коркина-Золотарева, так и независимо. Эксперимент показал 3-кратное ускорение работы блочного метода Коркина-Золотарева на четырехядерной системе. С использованием алгоритма были получены 1-е и 6-е места на конкурсах поиска коротких векторов.
\end{abstract}

\begin{Definition}  Базис $B=\{ b_{1} ,b_{2} ,\ldots ,b_{m} \} $ решетки $L\subset R^{n} $ приведен блочным методом Коркина-Золотарева(BKZ, Block Korkin-Zolotarev method, \cite{S1994}) блоком $\beta $, если:

\begin{enumerate}
\item  базис $B$ приведен по длине;

\item  $\left\| b_{i}^{\bot } \right\| =\lambda _{1} (L_{i} ),i=1,\ldots , m,$ где   $\lambda _{1} (L_{i} )$-кратчайший вектор в обратной (сопряженной) решетке $L_{i} $, образованной ортогональным дополнение пространства векторов $b_{i} ,\ldots ,b_{\min (i+\beta -1,m)} $.
\end{enumerate}
\end{Definition}

BKZ-метод содержит два основных алгоритма подлежащих распараллеливанию: ортогонализацию базиса решетки и поиск кратчайшего вектора. Вопрос распараллеливания ортогонализации базиса рассматривался в работе \cite{U2012}.  

Решение задачи поиска кратчайшего вектора будет заключаться, в полном переборе всех линейных комбинаций векторов базиса решетки $\left\| x\right\| ^{2} =\left\| \sum _{i=1}^{m}x_{i} b_{i}  \right\| ^{2} \le A^{2} ,x_{i} \in Z$, где  $A$ - норма искомого кратчайшего вектора. В качестве нормы берётся верхняя оценка длины кратчайшего вектора, $A=\sqrt{\gamma _{m} } \det (L)^{\frac{1}{m} } $,  где $\gamma _{m} $-константа Эрмита, в тех случаях, когда наименьший из векторов в базисе решетки превосходит оценку,  $\left\| b_{1} \right\| >\sqrt{\gamma _{m} } \det (L)^{\frac{1}{m} } $, \cite{HS2007}. По этой причине предварительное приведение базиса решетки позволяет уменьшить пространство перебора $x_{i} $. С целью уменьшения пространства перебора распишем базис решетки, через ортогональные вектора, для простоты изложения (сопряженной при вычислении на практике с утратой преимуществ параллелизма, характерных для современных QR-методов разложения) используя метод ортогонализации Грамма-Шмидта. Получим $b_{i} =\sum _{j=1}^{i}\mu _{i,j} b_{j}^{\bot }  ,2\le i\le m,1\le j<i\le m$, где $\mu _{i,j} $ - коэффициенты Грама-Шмидта. Легко убедиться, что в этом случае поиск кратчайшего вектора сводиться к решению системы неравенств: 

$\left\{\begin{array}{c} {x_{m}^{2} \left\| b_{m}^{\bot } \right\| ^{2} \le A^{2} ,} \\ {(x_{m-1} +\mu _{m,m-1} x_{m} )\left\| b_{m-1}^{\bot } \right\| ^{2} \le A^{2} -x_{m}^{2} \left\| b_{m}^{\bot } \right\| ^{2} } \\ {\ldots } \\ {(x_{1} +\sum _{i=2}^{m}x_{i} \mu _{i,j}  )^{2} \left\| b_{1}^{\bot } \right\| ^{2} \le A^{2} -\sum _{j=2}^{m}l_{j}  } \end{array},\right. $ где $l_{j} =(x_{j} +\sum _{i=j+1}^{m}x_{i}  \mu _{i,j} )^{2} \left\| b_{j}^{\bot } \right\| ^{2} $

и выбору одного из целочисленных векторов, у которого норма скалярного произведения с базисом решетки минимальна.

Формализуя задачу, получим обход дерева от корня к листу, в каждой из вершин которого решается соответствующее линейное уравнение.  Из корня этого дерева выходит $2\cdot \left\lceil \frac{A}{\left\| b_{m}^{\bot } \right\| } \right\rceil =2\cdot \left\lceil \frac{\sqrt{\gamma _{m} } \det (L)^{\frac{1}{m} } }{\left\| b_{m}^{\bot } \right\| } \right\rceil $ ветвей или $2\cdot \left\lceil \frac{\left\| b_{1} \right\| }{\left\| b_{m}^{\bot } \right\| } \right\rceil $, в случае предварительного приведения базиса решетки. В силу симметричности дерева (по свойствам нормы), для получения искомого кратчайшего вектора нам необходимо перебрать только половину его вершин. В результате полного обхода дерева от корня к листу, мы будем получать предполагаемый кратчайший вектор $x$  с нормой меньше либо равной искомой. Если норма полученного вектора будет меньше заданной ранее, целесообразно обновить ее, с целью уменьшения пространства перебора. Остановка алгоритма осуществляется, когда завершен обход вершин дерева или мы получили вектор с достаточной для нас нормой, в случае поиска короткого вектора. Каждый из потоков осуществляет вычисление своей ветки исходящей из корня дерева. 

\begin{Algorithm}[ht]
\caption{Поиска кратчайшего вектора в решетке $L(B),B=\{ b_{1} ,b_{2} ,\ldots b_{m} \} $.}
\label{algo}
\item[ \textbf{Вход:} $\left\| b_{1}^{\bot } \right\| ,\left\| b_{2}^{\bot } \right\| ,\ldots ,\left\| b_{m}^{\bot } \right\| $, $\mu _{i,j} $,  $id$- номер потока, начиная с 1]
\item[ \textbf{Выход:} Вектор $x=(x_{1} ,x_{2} ,...,x_{m} )\in Z^{m} :\left\| \sum _{i=1}^{m}x_{i} b_{i}  \right\| =\lambda _{1} (L(B))$]
\STATE $x=\{ 1,0,0,...,0\} ,l_{i} =\{ 0\} ^{m} ,X=\{ \emptyset \} ;$
\FOR {$i=1,...,m:$ $l_{i} =(x_{i} +\sum _{j=i+1}^{m}x_{j}  \mu _{j,i} )^{2} \left\| b_{i}^{\bot } \right\| ^{2} $,}
\STATE   \textbf{Если} $(\sum _{j=i+1}^{m}l_{j} >A ):i=i+1,x_{i} =x_{i} +1$; 
\STATE   \textbf{Если} $(\sum _{j=1}^{m}l_{j} \le A $ и $i=1)$: \textbf{Если} $\left\| \sum _{j=1}^{m}x_{j}  b_{j} \right\| <A:A=\left\| \sum _{j=1}^{m}x_{j}  b_{j} \right\| $,
\STATE   $X=X+\{ x\} ^{m} =X+\{ \sum _{j=1}^{m}x_{j}  b_{j} \} ,x_{1} =x_{1} +id$; \label{id}
\STATE   \textbf{Если} $(\sum _{j=i}^{m}l_{j} \le A $ и $i\ne 1):i=i-1,x_{i} =\left\lceil -\sum _{j=i+1}^{m}(x_{j} \mu _{ji} )-\frac{\sqrt{A-\sum _{j=i+1}^{m}l_{j}  } }{\left\| b_{i}^{\bot } \right\| }  \right\rceil $;
\ENDFOR
\PRINT вектор с минимальной нормой из множества $X$.  \label{ printing }
\end{Algorithm}

Данный алгоритм был реализован на основе потоковой модели NPTL (Native POSIX Thread Library, \cite{K2010}) под CentOS 6.3. Осуществляя приведение 103-мерной решетки BKZ-методом при $\beta =52$, 4-х потоках исполняемых на AMD Phenom 965/8 Gb DDR2-800, продемонстрировал 3-кратное ускорение выполнения метода по сравнению c fplll-4.0.1, \cite{fplll}.  С использованием данного алгоритма были получены 1-е и 6-е места на международном конкурсе алгоритмов поиска коротких векторов \cite{ILSVP}, для норм $A=m\cdot \det (L)^{\frac{1}{m} } $ и $A=1.05\cdot \frac{\Gamma \left(\frac{m}{2} +1\right)^{\frac{1}{m} } }{\sqrt{\pi } } \cdot \det (L)^{\frac{1}{m} }$, соответственно.

\enabstract{
This article present a parallel CPU implementation of Kannan algorithm for solving shortest vector problem in Block Korkin-Zolotarev lattice reduction method. Implementation based on Native POSIX Thread Library and show linear decrease of runtime from number of threads.
\protect\enkeywords{shortest vector problem, SVP, block Korkin-Zolotarev, BKZ, lattices, parallel algorithms.}
}

\begin{authors}
    \item{Усатюк Василий Станиславович}{программист кафедры дискретной математики и защиты информации Братского государственного университета, г. Братск}{L@Lcrypto.com}
\end{authors}

\end{document}